\documentclass[preprint,amsmath,amssymb,aps,prb
]{revtex4-1}
\pdfoutput=1
\usepackage{graphicx}
\usepackage{epsfig}
\usepackage{subfigure}
\usepackage{dcolumn}
\usepackage{bm}
\usepackage[mathlines]{lineno}
\begin{document}

\preprint{APS/123-QED}

\title{Anomalous increasing of permeability peak value of optical mode ferromagnetic resonance in the exchange coupled bilayer system}

\author{Wenfeng Wang}

\author{Lulu Pan, Wenjie Song}%

\author{Guozhi Chai}
 \email{chaigzh@lzu.edu.cn}

\author{Desheng Xue}
 \email{xueds@lzu.edu.cn}
\affiliation{ Key Laboratory for Magnetism and Magnetic Materials of the Ministry of Education, Lanzhou University,
 Lanzhou,730000, People¡¯s Republic of China.}%
\date{\today}

\begin{abstract}
Acoustic and optical ferromagnetic resonance (FMR) in the interlayer exchange coupled Fe$_{20}$Ni$_{80}$/Co bilayer have been investigated. In the optical mode, unexpected increasing tendencies of peak value at the resonance frequency has been observed under an increasing magnetic field. We presented analytical calculations with which the exchange coupling between Co and Fe$_{20}$Ni$_{80}$ layers, the magnetization and the in-plane uniaxial anisotropy are taken into account, to interpret the increasing of the maximum values of the optical permeability. Both experimental measurements and theoretical calculation show that such tendencies are dependent on the layer thickness $t$, and that there is a critical field above which the optical peak value begins to decrease. These results might help us to understand the mechanism of interlayer exchange coupling induced optical FMR and might enlighten us to find new possibility of high frequency applications of magnetic materials.
\end{abstract}
\pacs{Valid PACS appear here}
\maketitle


\section {\label{sec:section1}Introduction}
For decades, the magnetic materials possessing excellent soft magnetic properties are most wanted in many electronic devices, yet it has been a challenge to alter the magnetic properties of a material once it was fabricated. From the points of view of practical use of magnetic properties, the dynamic properties of a magnetic material are governed by the complex dynamic permeability\cite{Gilbert2004,Zhang2009}: $\mu=\mu^{'}-i\mu^{''}$, where $\mu^{'}$ denotes the real part of permeability, which determines the reorientation angle of the magnetization processional mode, whereas the imaginary permeability $\mu^{''}$ determines the energy dissipation\cite{Jiang2018}. Materials with large $\mu^{'}$ and small $\mu^{''}$ implicate great performance and small energy consumption, respectively. Therefore, a magnetic material with large and controllable $\mu$ at relative high work frequency range is desperately demanded. However, both real and imaginary permeability of the magnetic materials with traditional magnetization dynamics will reduce to the extent of useless owing to the restriction of Acher's limit\cite{Acher2000,Chai2008}. From this point of view, one can have only a descending of $\mu$ with the increase of the resonance frequency. Consequently, the fulfillment of a magnetic material, of which $\mu$ increase with increasing frequency, is therefore very desirable for modern electronic devices.

To meet this demand, new kinds of magnetic material structure consisting of two or more  magnetic layers are then carried out by several authors. In these multilayered systems, two modes of FMR, known as acoustic and optical FMR, respectively, can be observed in experiments\cite{Pomerantz1987,Krebs1990}, owing to the exist of so called interlayer exchange coupling. The energy of the interaction are described by two exchange coupling parameters, the bilinear coupling $J_1$\cite{Hoffmann1970} and the biquadratic coupling $J_2$\cite{Slonczewski1991,Slonczewski1995}. Generally, $J_1$ is dominant in the films, where moments of the two layers are parallel\cite{Baselgia1988,Layadi1990,Guslienko1991,Wigen1992,Lindner2003} or antiparallel\cite{Grunberg1986,Krebs1989,Heinrich1990} to each other to meet the demand of energy minimization. While some authors have shown that the biquadratic coupling can also become dominating\cite{Grunberg2000,Layadi2002}, leading to a 90$^{\circ}$-type coupling, in which the moments of the two ferromagnetic layers are vertical to each other. It was found that the biquadratic exchange coupling arise from spatial fluctuation of the interlayer thickness in a sandwiched structure\cite{Slonczewski1991}.

For the $J_1$ exchange coupling (ferromagnetic or antiferromagnetic corresponding to $J_1$$>$0 or $J_1$$<$0) dominated cases, the exchange energy per unit area at the interface can be written as,
\begin{equation}
E_{ex}=-J\frac{\mathbf{M_1}\cdot \mathbf{M_2}}{M_1M_2}
\label{eq:Eex},
\end{equation}

where $J$ is the bilinear exchange coupling coefficient with the unit of erg/cm$^2$, $M_1$ and $M_2$ are the saturation magnetizations of the individual layers. Note that we neglected the subscript of $J$ here for simplicity. It has been found that the exchange coupling strength depend dramatically on the thicknesses of the magnetic films\cite{Pomerantz1987,Layadi1990}, as well as the thickness of the interlayers\cite{Grunberg1986}. Heinrich and coworkers found out that the magnetic coupling in epitaxial bcc Fe(001)/Cu(001) /Fe(001) trilayers changes from ferromagnetic to antiferromagnetic as the Cu(001) interlayer thickness changes. The exchange energy can be derived from frequency difference of the acoustic and optical modes\cite{Chen2015b,LiShD2016}. For the ferromagnetic coupling, the resonance frequency of optical mode is higher than that of acoustic mode, while for the antiferromagnetic coupling, the optical resonance frequency is lower than that of acoustic mode. The dispersive relations and dependence of resonance intensities on exchange coupling field, saturation magnetization and anisotropic field can be obtained from solving Landau-Lifshitz-Gilbert equeations\cite{Heinrich1988,Layadi1990,Wigen1992,Layadi2001,Layadi2005}. Note that the resonance intensity mentioned here is defined as relating to the area under FMR absorption line.

In experimental, FMR is one of the most ubiquitous technique to investigate the micro wave magnetic properties of interlayer exchange coupled magnetic multi-layers\cite{Kittel1948,Smit1955a,Heinrich1987,Farle1999,Layadi2016}. By measuring the permeability spectra as a function of exchange coupling strength, magnetization, anisotropy, damping factor, magnetic layer thickness and even applied magnetic field, one can obtain a lot of basic magnetic properties of the multi-layer systems, and can test theories that describe the mechanisms of these coupling structures.

In this paper, we implemented a theoretical calculation and experimental measurements of permeability for a exchange coupled bilayer system consisting of two different ferromagnetic layers in intimate contact. Both numerical calculation and FMR results for the permeability of the bilayer are presented. Differences of acoustic and optical resonance behaviors under a increasing external magnetic field $\bf{H}$ are studied with several parameters are taken into account including layer thickness $t$, magnetization $M$, in-plane uniaxial anisotropy (IPUMA) $\bf{H}_k$, damping factor $\alpha$, and exchange strenth $J$.

\section{experimental details}
\begin{figure*}
\includegraphics[width=12 cm]{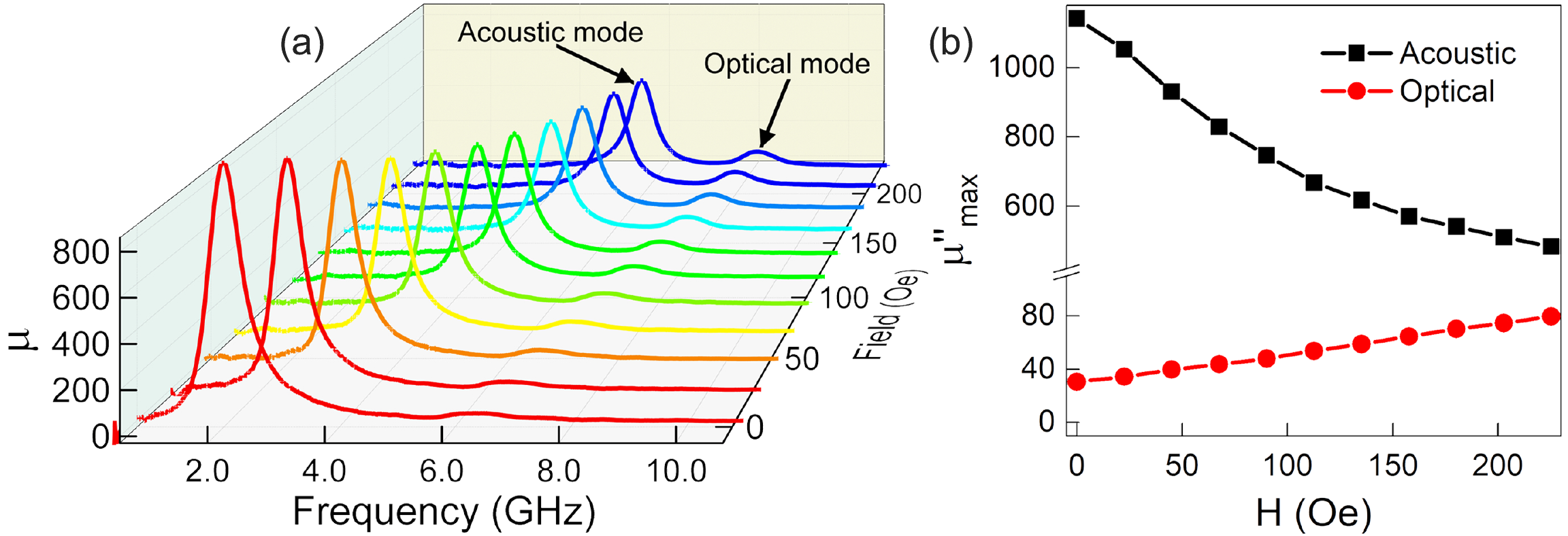}
\caption{\label{fig:field_perm}(a) The imaginary permeability of Fe$_{20}$Ni$_{80}$/Co bilayer measured with a increasing magnetic field applied along the easy direction, (b) the peak values of acoustic (black) and optical (red) imaginary permeability of the Fe$_{20}$Ni$_{80}$/Co bilayer as functions of applied field .}
\end{figure*}
A set of Fe$_{20}$Ni$_{80}$/Co bilayer films with different layer thicknesses are fabricated and studied in this paper. The samples were grown by radio frequency (rf) magnetron sputter deposition on 0.43 mm thick Si (111) substrates, which were attached to oblique sample holders with oblique angle of $30^\circ$ to induce the in-plane uniaxial anisotropy, in a ultrahigh vacuum chamber. The base pressure of the chamber prior to sputtering was pumped to approximately $6\times10^{-5}$ Pa. The deposition pressure during the fabrication was maintained at 0.3 Pa at Ar ambient with gas rate flow of 10 SCCM (cubic centimetre per minute at STP). The sputter targets of Co metal and Fe$_{20}$Ni$_{80}$ alloy are 3 inches in diameter. The rf power of 50 W was used to deposit the films. The thicknesses of the films were controlled by controlling the deposition time for each layer. The schematic of the film structure is shown in Fig.~\ref{fig:sketch}, where FM1 represents Co layer and FM2 is Fe$_{20}$Ni$_{80}$ layer in this case. For simplicity, the Co layer thickness $d_1$ was fixed to be 28.6 nm, while the Fe$_{20}$Ni$_{80}$ layer have varying thicknesses, $d_2=33.4,~48.8$ and 66.8 nm.

The FMR measurements of the films were performed via vector network analyser (VNA, Agilent E8363B, USA) with a home made shorted-circuited microstrip line (MSL) jig connected to it through a subminiature assembly coaxial connector\cite{Liu2005,Wei2014a}. The resistance of the MSL is $50~\Omega$ to meet impedance matching of VNA's test port. During the measurements, the micro magnetic field $h$ was perpendicular to the easy axis (EA) of the samples. For each sample, an increasing planer magnetic field $\bf{H}$ along EA was applied.

The experimental results of dc magnetic field dependent imaginary permeability of the Fe$_{20}$Ni$_{80}$/Co bilayer are presented in Fig.~\ref{fig:field_perm} (a), the applied field ($H$)lies in the film plane and along the easy axis of the film. Compared to that of the acoustic mode, the resoannce frequency of the optical mode is higher, however the peak value of the imaginary permeability is significantly small. This is because that in the case of strong ferromagnetic coupling, the moments of the Co and Fe$_{20}$Ni$_{80}$ layers precess out-of-phase in optical FMR mode causing an offset of radio-frequency (rf) components of $\bf{M}_i$. Hence the peak value of the optical imaginary permeability of the bilayer is vastly crippled. A evident downward trend of the acoustic mode was observed whereas the absorption peak of the optical mode increased unusually with the increase of the applied field, which is \emph{completely opposite} to that of the acoustic mode. The difference of peak values of the two modes decreased from 1098 to 370 when $H$ increases from 0 to 225 Oe. This abnormal increase of optical peak value might have bearing on the out-of-phase precession of magnetic moments. The variation of peak values of the acoustic and optical imaginary permeability versus the applied field $H$ are presented in Fig.~\ref{fig:field_perm} (b), one can see that the two modes have completely opposite variation trend. In the following section, we proposed a theoretical model aimed at numerically calculating the permeability of the Fe$_{20}$Ni$_{80}$/Co bilayers to give a comprehensive understanding on the eccentric behaviour of optical resonance.

\section {\label{sec:section2}theoretical model}
\begin{figure}
\includegraphics[width=5 cm]{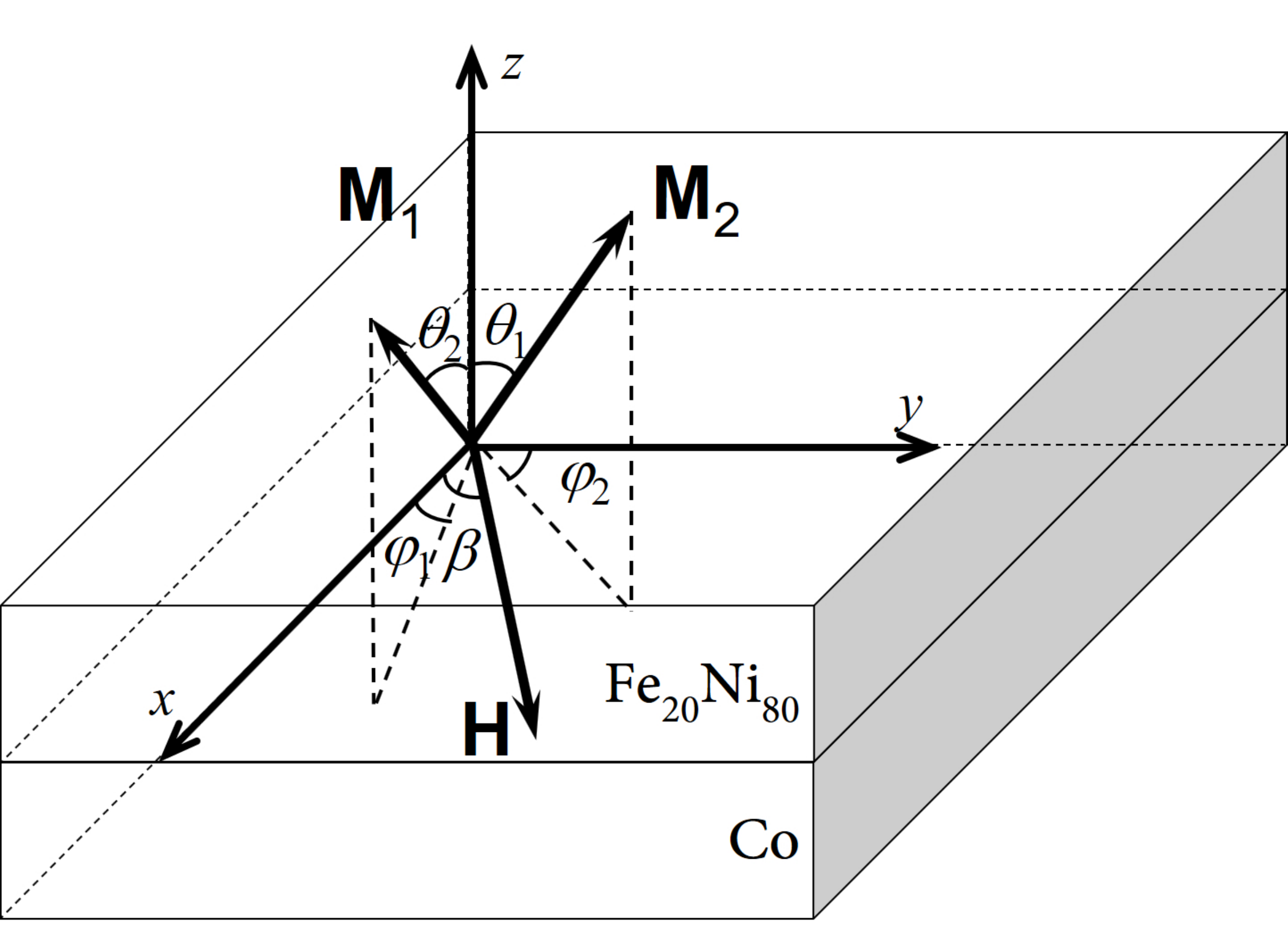}
\caption{\label{fig:sketch}Schematic show of the magnetic bilayer structure and coordinate systems. The axis $x$ is chosen to coincide with the uniaxial anisotropy of the sample, and the axis $y$ to be along the direction of micro magnetic field $h$.}
\end{figure}

The schematic of the bilayer structure is shown in Fig.\ref{fig:sketch}. Considering a bilayer system consisting of FM1 and FM2 layers lies in the $x-y$ plane with the axis $z$ normal to the film planes. The calculation is based on the LLG equation that the moments of the bilayer deviation from the equilibrium positions, with a microwave magnetic field perpendicular to the magnetization in the film plane. Only the situation of the external magnetic field $\bf{H}$ lying in the film plane, at an angle $\beta$ with respect to $x$ axis, is taken into consideration. The magnetization  $\bf{M}_i$ of FM$i$, is characterized by the angles $\theta_i$ and $\varphi_i$, where $i$ ($i$= 1, 2) denotes the FM1 and FM2 respectively. To simplify the calculations, we assume that the magnetization and uniaxial anisotropy of the two ferromagnetic layers are all lie in the film planes and that $\beta=0$, and that the biquadratic coupling is negligible compared to bilinear coupling, i.e., $J_1$${\gg}$$J_2$. Note that no magnetocrystalline anisotropies are considered for both layers.

With all these assumption above, the total free energy per unit of the system can be written as
\begin{eqnarray}
E=&&t_1[-M_1Hx_1+2\pi M_1^2z_1^2+K_1(y_1^2+z_1^2)]\nonumber\\
&&+t_2[-M_2Hx_2+2\pi M_2^2z_2^2+K_2(y_2^2+z_2^2)]\nonumber\\
&&-J(x_1x_2+y_1y_2+z_1z_2)
\label{eq:energy},
\end{eqnarray}
where $t_i$ and $K_i$ are the thickness and uniaxial anisotropy constant of FM$i$ ($i=1,2$), respectively. $H$ is the applied external magnetic field and $J$ is the exchange coupling constant. Note that the biquadratic coupling is not considered here, thus the bilinear coupling $J_1$ is written to $J$. $x_i$, $y_i$ and $z_i$ are the direction cosines of the $M_i$ to the $x$, $y$ and $z$ axes, respectively. They are give by
\begin{eqnarray}
x_1=&&\sin{\theta_1}\cos{\varphi_1}, ~y_1=\sin{\theta_1}\sin{\varphi_1}\nonumber\\
x_2=&&\sin{\theta_2}\cos{\varphi_2},~ y_2=\sin{\theta_2}\sin{\varphi_2}\nonumber\\
z_1=&&\cos{\theta_1},~z_2=\cos{\theta_2}
\label{eq:direction_cosine}.
\end{eqnarray}
The total energy $E$ consists of the Zeeman energy, the in-plane uniaxial anisotropy\cite{Layadi1987,Layadi2001} and the dipolar energy\cite{Rezende1998a} of FM1 and FM2 layers, as well as the exchange coupling energy between FM1 and FM2.

At equilibrium, the first derivatives of $E$ with respect to $x_i$, $y_i$ and $z_i$ must be equal to zero. It is apparent that when $x_i=1$ and $y_i=z_i=0$, the system reach to equilibrium condition, at which the magnetizations,  $\bf{M}_1$ and $\bf{M}_2$, are all lie in the $x$ direction in the films plane. Note that in the antiferromagnetic situation, $\bf{M}_1$ and $\bf{M}_2$ are antiparallel, i.e., $x_1\cdot{x_2}=-1$.

Now we consider the dynamic behaviors of $M_1$ and $M_2$ in a weak microwave magnetic field $\bf{h}$. It is described by Landau-Lifshitz equation with the Gilbert damping term
\begin{equation}
\frac{\mathrm{d}{\mathbf{M}_i}}{\mathrm{d}t}=\gamma{\mathbf{M}_i}\times[\nabla_{M_i}(\frac{E}{t_i})+\frac{\alpha}{\gamma M_i}\frac{\mathrm{d}{\mathbf{M}_i}}{\mathrm{d}t}-\mathbf{h}_i\mathrm{e}^{j\omega t}]
\label{eq:LLG}.
\end{equation}
Here $\gamma$ is the gyromagnetic ratio, $\alpha$ is the damping factor and $\omega$ is the angular frequency of $h_i$. Considering that the magnetization vectors excited by $h$ oscillate about the equilibrium position, the Eq.~(\ref{eq:LLG}) can be linearized by expanding the free energy $E$ in Taylor series up to second order. The motion of the
moments then can be written in matrix form as
\begin{small}
\begin{widetext}
\begin{eqnarray}
&&\left(
  \!\begin{array}{cccc}
    \gamma E_{yy}-j\alpha \omega M_1 &\gamma E_{yz}-j\omega M_1 & \gamma E_{ya} & \gamma E_{yb}\\
    \gamma E_{zy}+j\omega M_1 &\gamma E_{zz}+j\alpha \omega M_1 & \gamma E_{za} & \gamma E_{zb}\\
    \gamma E_{ay}&\gamma E_{az}& \gamma E_{aa}-j\alpha \omega M_1 & \gamma E_{ab}-j\omega M_1\\
    -\gamma E_{by}&-\gamma E_{bz}& \gamma E_{ba}+j\omega M_1 & \gamma E_{bb}+j\alpha \omega M_1 \\
  \end{array}\!
\right)\!\times\!\left(
               \begin{array}{c}
                \Delta{y}\\
                \Delta{z}\\
                \Delta{a}\\
                \Delta{b}\\
               \end{array}
             \right)\!=\!\left(
               \!\begin{array}{c}
                \gamma M_1h_y\\
                0\\
                \gamma M_2h_y\\
                0\\
               \end{array}\!
             \right)
\label{eq:matrix1}
\end{eqnarray}
\end{widetext}
\end{small}
where $E_{ij}=\partial^2E/{\partial i \partial j}$ are the second partial derivative of energy with respect to $i$ and $j$ $(i,j=y,z,a,b)$ at the equilibrium position.
Note that, in Eq.~(\ref{eq:matrix1}), we let $y_1=y,~z_1=z,~y_2=a$ and $z_2=b$ to make the calculation simple, and $\Delta y,~\Delta z,~\Delta a$ and $\Delta b$ denote the small variations of $y,~z,~a$ and $b$, respectively. By substituting Eq.~(\ref{eq:energy}) into Eq.~(\ref{eq:matrix1}) and simplifying the matrix equation, we can obtain
\begin{widetext}
\begin{eqnarray}
&&\left(
  \begin{array}{cccc}
   \Omega_{k1}-j\alpha \omega &-j\omega &-\omega_{J2} &0\\
    j\omega &\Omega_{m1}+j\alpha \omega & 0 & -\omega_{J2}\\
    -\omega_{J1}& 0 & \Omega_{k2}-j\alpha \omega & \-j\omega \\
    0&-\omega_{J1}& j\omega & \Omega_{m2}+j\alpha \omega \\
  \end{array}
\right)\times\left(
               \begin{array}{c}
                \Delta{m_{1y}}\\
                \Delta{m_{1z}}\\
                \Delta{m_{2y}}\\
                \Delta{m_{2z}}\\
               \end{array}
             \right)=\left(
               \begin{array}{c}
                \omega_{m1}h_y\\
                0\\
                \omega_{m2}h_y\\
                0\\
               \end{array}
             \right)
\label{eq:matrix2}
\end{eqnarray}
\end{widetext}
with
\[\Omega_{k1}=\omega_0+\omega_{k1}+\omega_{J1},~\Omega_{m1}=\omega_0+\omega_{k1}+\omega_{m1}+\omega_{J1}\]
\[\Omega_{k2}=\omega_0+\omega_{k2}+\omega_{J2},~\Omega_{m2}=\omega_0+\omega_{k2}+\omega_{m2}+\omega_{J2}\]
where $\omega_0=\gamma H,~\omega_{ki}=\gamma\dfrac{2K_i}{M_i},~\omega_{Ji}=\dfrac{\gamma J}{d_iM_i}$ and $\omega_{mi}=4\pi\gamma M_i (i=1,2)$. On solving this equation with the help of computer, one can obtain the real and imaginary permeability of both acoustic and optical modes for the bilayer systems numerically.

\section{Results and discussions}
\begin{figure}
\includegraphics[width=8.4 cm]{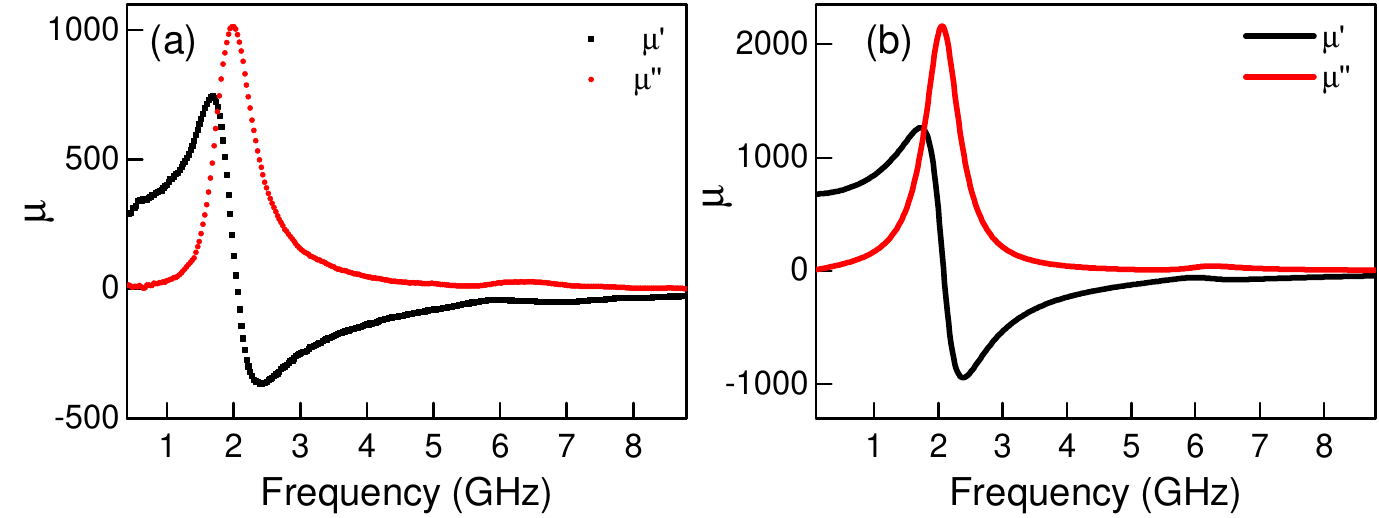}
\caption{\label{fig:spectra}(a) Permeability of Fe$_{20}$Ni$_{80}$/Co bilayer measured at zero external dc magnetic field. (b) Numerical results of zero-field permeability of the bilayer.}
\end{figure}
Fig.~\ref{fig:spectra} (a) shows the permeability spectra of the Fe$_{20}$Ni$_{80}$/Co bilayer at zero external dc magnetic field, with the thicknesses of Co and Fe$_{20}$Ni$_{80}$ layer are $t_1=28.6$ nm and $t_2=48.8$ nm, respectively. Two FMR modes are observed in the permeability spectra, one is the acoustic mode ($f_{ac}=1.99$ GHz), the other one is the optical mode ($f_{op}=6.31$ GHz). Numerical results of zero-field permeability of the Fe$_{20}$Ni$_{80}$/Co bilayer according to Eq.~(\ref{eq:matrix2}) are presented in Fig.~\ref{fig:spectra} (b). The parameters used in the calculation are $4\pi M_1=15.2$ kG, $H_{k1}=44$ Oe, $\alpha_1=0.028$ for Co layer; $4\pi M_2=11.9$ kG, $H_{k2}=36$ Oe, $\alpha_2=0.008$ for Fe$_{20}$Ni$_{80}$ layer, and the interlayer exchange coupling strength is $J=0.54$ erg/cm$^2$, they are in accordance with experimental measurements. A large resonance frequency of $f_{op}=6.31$ GHz is obtained in the optical mode owing to the presence of the interlayer exchange coupling.

\begin{figure}
\includegraphics[width=6 cm]{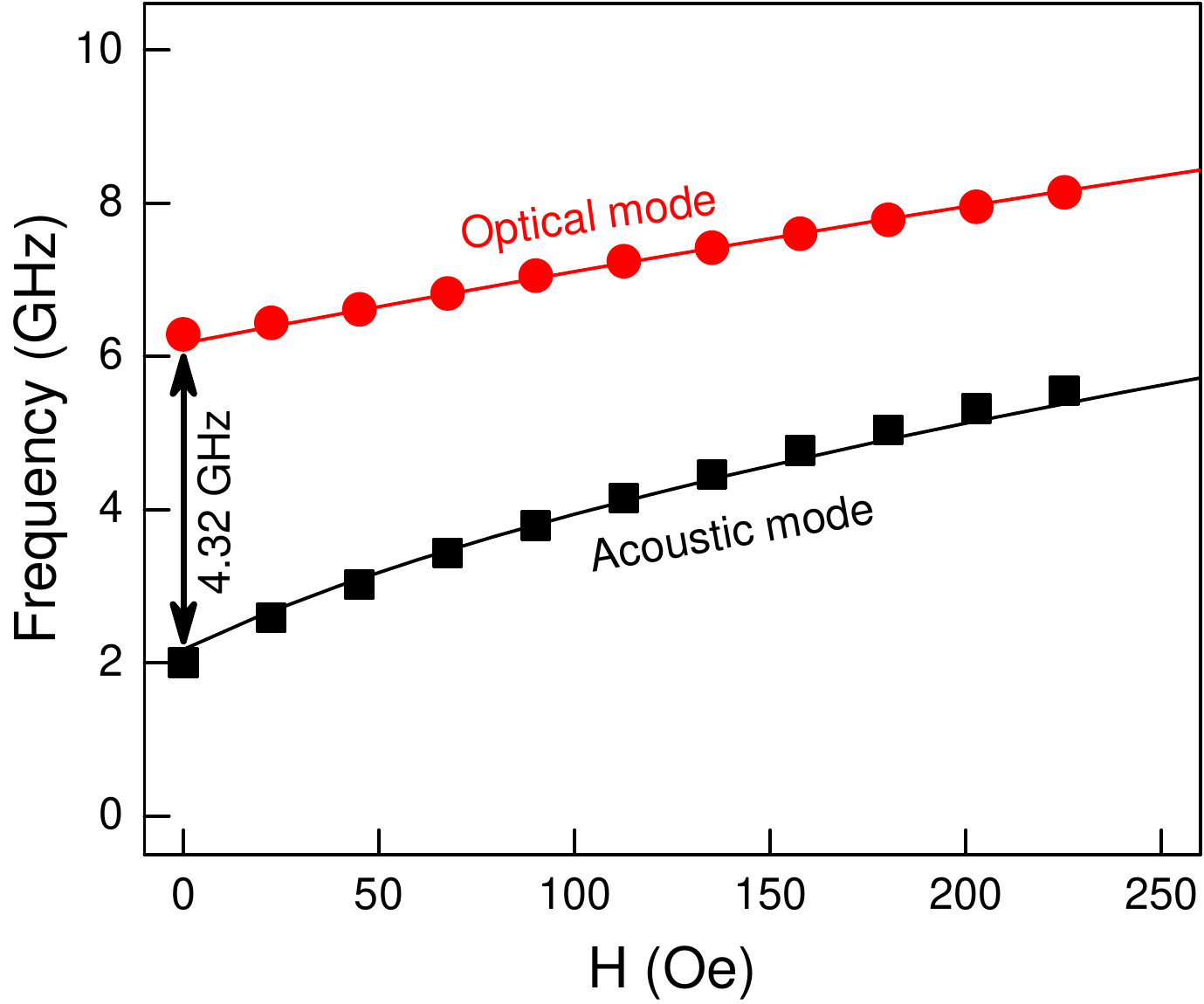}
\caption{\label{fig:freq_field}
The resonance frequencies vs applied magnetic field $H$ of the Fe$_{20}$Ni$_{80}$/Co bilayer, where the solid black square and circle represent the experimental data and the solid lines denote the calculated dispersion relation.}
\end{figure}
Fig.~\ref{fig:freq_field} show the resonance frequencies of the Fe$_{20}$Ni$_{80}$/Co bilayer as the functions of applied magnetic field. The resonance frequency of the optical mode at zero field is 4.32 GHz higher than that of the acoustic mode. For the acoustic mode, the magnetization vectors of Co and Fe$_{20}$Ni$_{80}$ layers precess in phase and the dispersive relation of acoustic mode degenerate with that of single layer because that the interlayer exchange coupling produces no dynamic contributions to the resonance. For the optical mode, however, the moments of the two layers precess out of phase and therefore the coupling produces effect exchange fields of $H_{e1}=J/d_1M_1=139$ Oe and $H_{e2}=J/d_2M_2=149$ Oe in Co and Fe$_{20}$Ni$_{80}$ layers, respectively\cite{Zhang1994,Zhang1994a}. The resonance frequencies of both acoustic and optical modes increase as $H$ increased, the black and red line in Fig.~\ref{fig:freq_field} denote the calculated dispersive relations of the two FMR modes, respectively, which are in accordance with the experimental measurements.

\begin{figure*}
\includegraphics[width=10 cm]{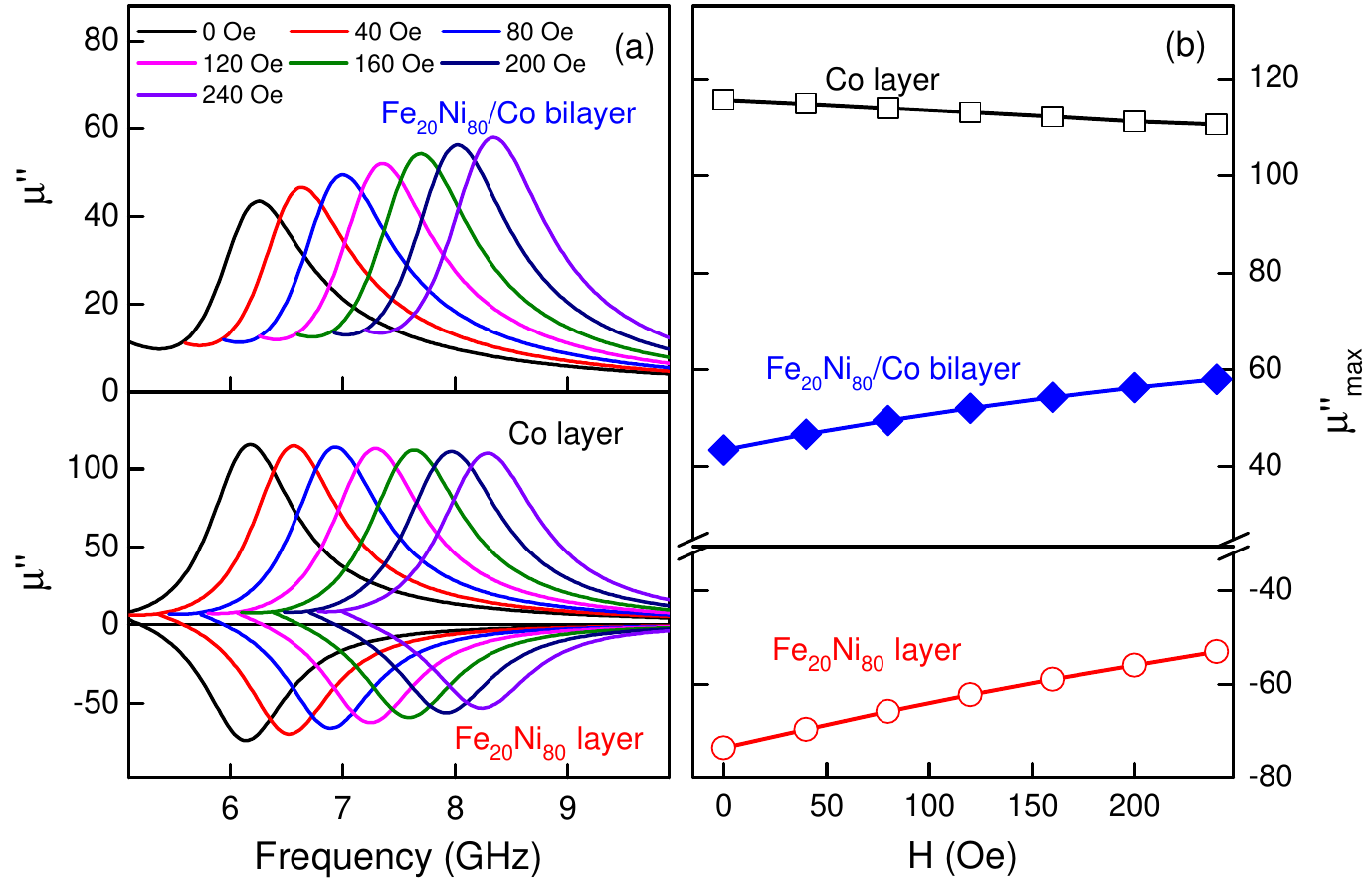}
\caption{\label{fig:perm_simu}(a) The calculated field-dependent optical imaginary permeability of Fe$_{20}$Ni$_{80}$/Co bilayer, as well as separated Co and Fe$_{20}$Ni$_{80}$ layers respectively. (b) The absorption peak values dependence on applied magnetic field. The parameters are the same with the Fig.~\ref{fig:spectra} (b)}
\end{figure*}
As previously presented in Fig.~\ref{fig:field_perm}, the peak value of optical FMR varies abnormally with respect to $H$, which is different from that of acoustic mode. In Fig.~\ref{fig:perm_simu} (a), we show the numerically calculated optical imaginary permeability spectra of the Fe$_{20}$Ni$_{80}$/Co bilayer and of the separated Co and Fe$_{20}$Ni$_{80}$ layers, respectively, with applied field $H$ increase from 0 to 240 Oe, to understand the abnormal field-dependent behavior of the optical resonance. The optical permeability of Co and Fe$_{20}$Ni$_{80}$ layers have different signs [See Fig.~\ref{fig:perm_simu} (a)]corresponding to the in-phase and out-of phase precession of the moments, respectively. One can see that the absolute peak values of separated Co and Fe$_{20}$Ni$_{80}$ layers decay at different speeds with the increase of $H$, leading to an increase of integral peak value of the bilayer. The peak values of  the bilayer, as shown in Fig.~\ref{fig:perm_simu} (b), equals numerically to the sum of those of separated Co and Fe$_{20}$Ni$_{80}$ layers, suggesting that the net rf component of the magnetization vectors in the bilayer comes from those of vector superpositions of Fe$_{20}$Ni$_{80}$ and Co layers.

\begin{figure*}
\includegraphics[width=13 cm]{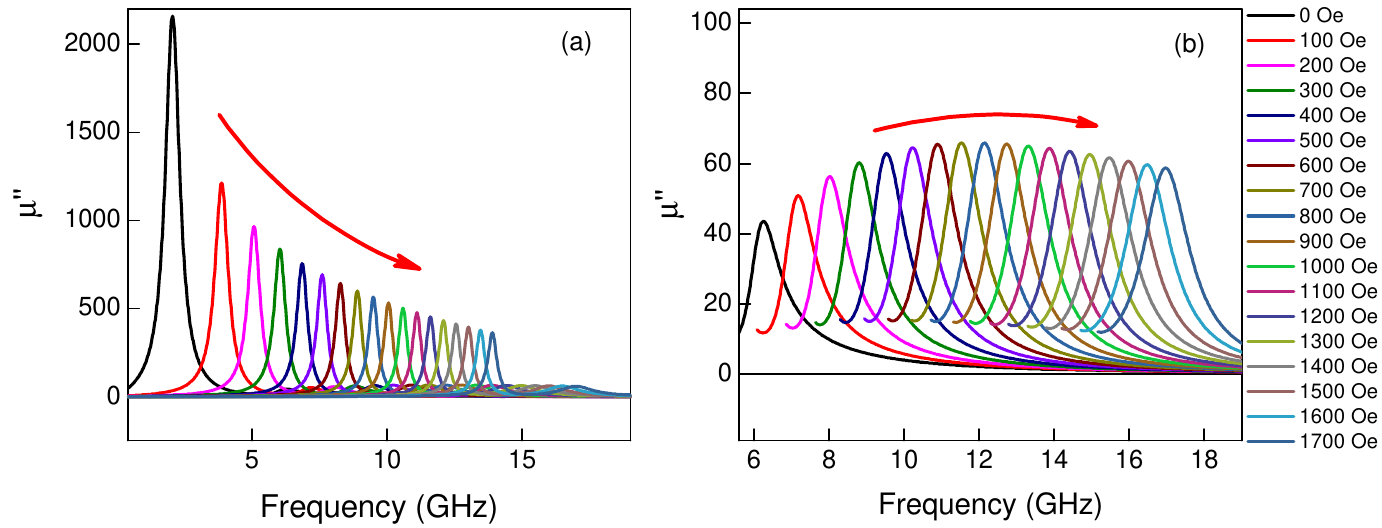}
\caption{\label{fig:large_field}The calculated acoustic (a) and optical (b) imaginary permeability of Fe$_{20}$Ni$_{80}$/Co bilayer with the applied field increasing from 50 to 1650 Oe. The simulation parameters are the same with the Fig.~\ref{fig:spectra} (b).}
\end{figure*}
Due to the presence of uniaxial anisotropy $\bf{H}_k$ and applied field $\bf{H}$, when applying a transverse microwave magnetic field $h$, the transverse (along the $\bf{h}$ directions) rf components of the magnetization of Co and Fe$_{20}$Ni$_{80}$ layers are always antiparallel to each other resulting in a small resonance absorption in optical mode. When $\bf{H}$ is applied along the easy axis, the peak value of acoustic mode decreases with the increase of $H$ owing to reduction of rf components of the two layers. While that of the optical mode increases as $H$ increases, this is related to the different reduction speed of rf components of the magnetization vectors in two layers.

In Fig.~\ref{fig:large_field}, we plot the simulated imaginary permeability of the Fe$_{20}$Ni$_{80}$/Co bilayer, the thickness of Fe$_{20}$Ni$_{80}$ is 48.8 nm, the applied field $H$ increases from 0 to 1700 Oe. A significantly downward tendency of the peak value is observed in the acoustic mode when $H$ increases as shown in Fig.~\ref{fig:large_field} (a). For the optical mode, the variation trajectory of peak value of the imaginary permeability is divided into two parts. Let us define a critical field $H_{crt}$, below which the peak value increases gradually as $H$ increases while above which the peak value begins to decrease. In this case, the $H_{crt}$ is about 750 Oe, as seen in Fig.~\ref{fig:large_field} (b), the peak value reaches its maximum at this point. This is due to the vector superposition effect of the rf components of the magnetization vectors in two ferromagnetic layers that result in an increase of optical mode at a small field. However, when $H>H_{crt}$, the magnetization vectors of both layers are completely saturated along the field direction leading to small precession angles, which means the value of the optical resonance peak begins to decrease with respect to $H$.

\begin{figure}
\includegraphics[width=7 cm]{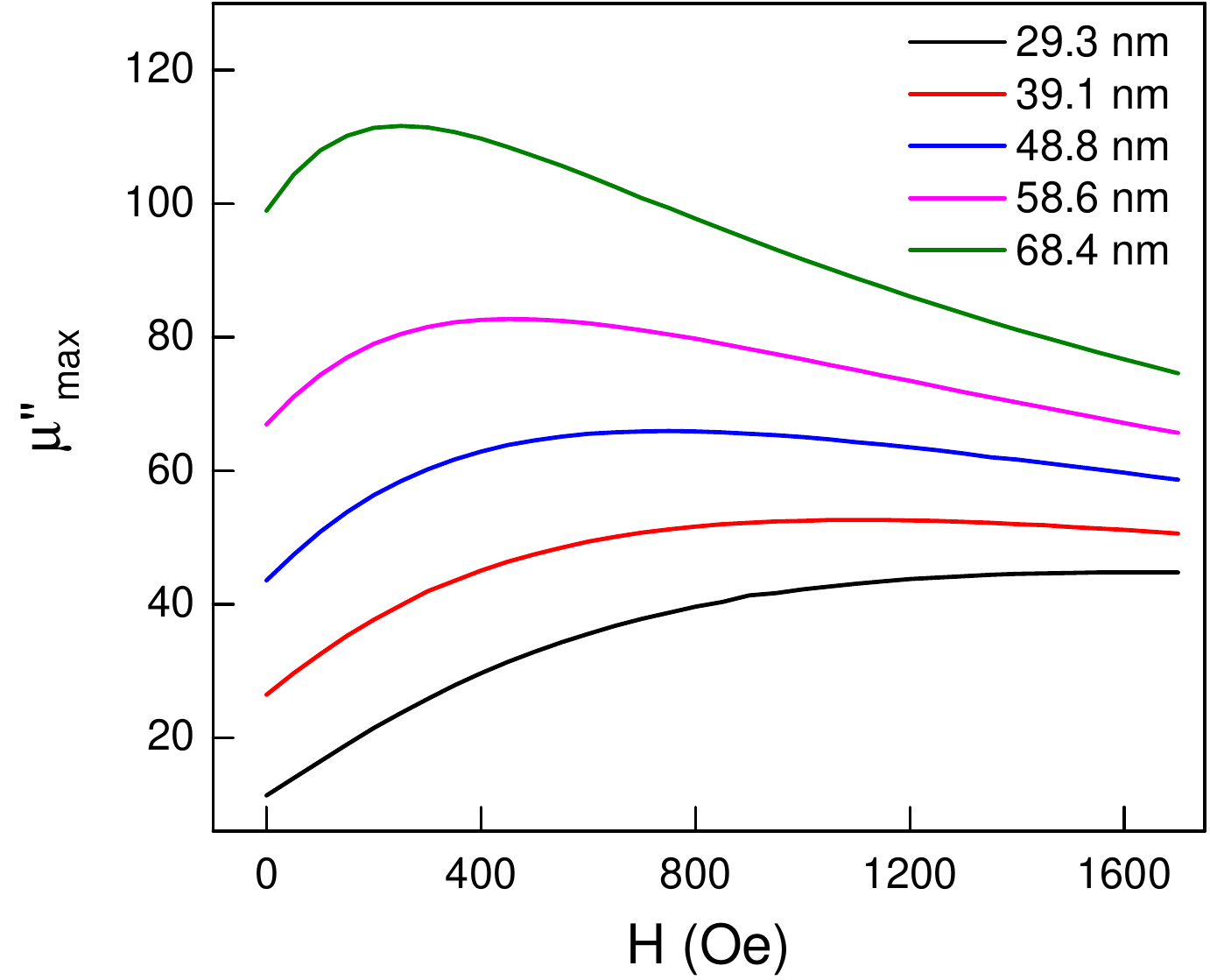}
\caption{\label{fig:maximum_field}Calculated peak value of the optical imaginary permeability with respect to $H$ for the Fe$_{20}$Ni$_{80}$/Co bilayers with Fe$_{20}$Ni$_{80}$ layer thickness increasing from 33.4 nm to 66.8 nm.}
\end{figure}
In previous sections we have already discussed that the optical FMR originates from the phase difference of the magnetic moments precession between Fe$_{20}$Ni$_{80}$ and Co layers. The interlayer exchange coupling plays a key role in this process, which can be adjusted by changing the layer thickness\cite{Zhang1994,Zhang1994a}. Thus one can consequently put in mind that there is a connection between layer thickness and optical FMR. Fig.~\ref{fig:maximum_field} shows the optical imaginary permeability peak values versus $H$ for a series of Fe$_{20}$Ni$_{80}$/Co bilayers with Fe$_{20}$Ni$_{80}$ layer thickness $t_2$ increasing from 33.4 to 66.8 nm. The Co layer thickness remains unchanged. The measurements of the permeability spectra at higher frequency are limited by our home made shorted-circuited MSL jig, the permeability spectra with applied field $H$ larger than 250 Oe are then exceed the test range, therefore, only numerically simulated results are shown here. From Fig.~\ref{fig:maximum_field}, one can see how the Fe$_{20}$Ni$_{80}$ layer thicknesses affect the variation tendency of the optical mode peak value. For the samples with Fe$_{20}$Ni$_{80}$ layer thinner than 48.8 nm, the peak values change slowly but with high $H_{crt}$, when Fe$_{20}$Ni$_{80}$ layer grown thicker, $H_{crt}$ is reduced and more dramatic changes of the peak values are observed.

\begin{figure}
\includegraphics[width=7 cm]{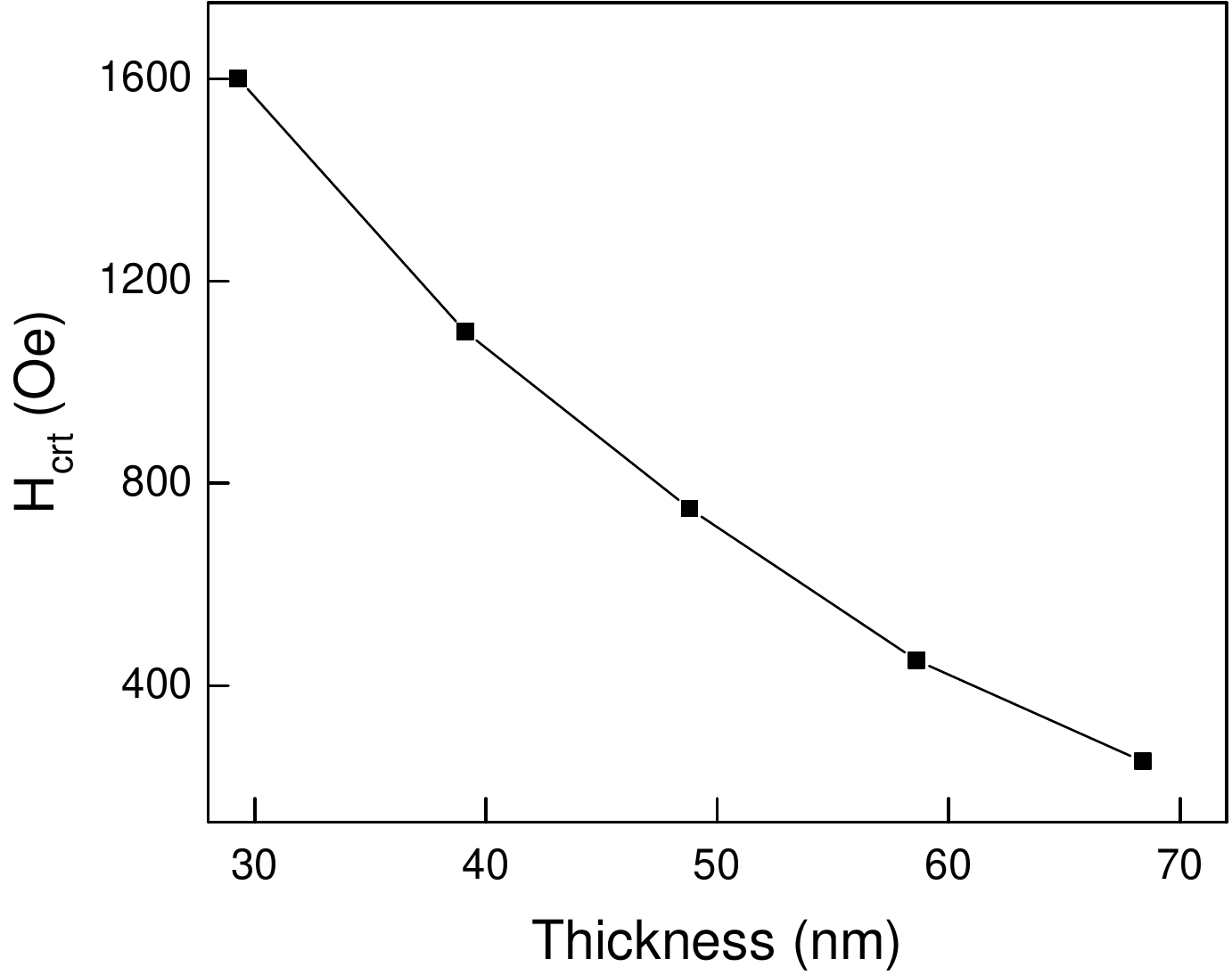}
\caption{\label{fig:maximum_thick}The critical field $H_{crt}$ vs Fe$_{20}$Ni$_{80}$ layer thickness.}
\end{figure}
Fig.~\ref{fig:maximum_thick} plots the critical field $H_{crt}$ changes with respect to Fe$_{20}$Ni$_{80}$ layer thickness. $H_{crt}$ can be vastly controlled through adjusting the Fe$_{20}$Ni$_{80}$ layer thickness of Fe$_{20}$Ni$_{80}$/Co bilayers, a possible explanation is that with Fe$_{20}$Ni$_{80}$ layer thickness increases, the equivalent exchange coupling fields $H_{e1}=J/d_1M_1$ and $H_{e2}=J/d_2M_2$ are reduced, which might have influence on $H_{crt}$.
\section{conclusion}
In summery, a theoretical model describing dynamic behaviors of the magnetic moments of an exchange coupled bilayer system under a microwave magnetic field was developed. A comparison of calculated permeability and the experimental data was investigated, and found that for a ferromagnetic coupled bilayer, the FMR frequency of optical mode is much higher than that of acoustic mode, corresponding to an exchange strength of $J=0.54$ erg/cm$^2$. The numerical simulations are in well accordance with the experimental results. The effect exchange coupling field for the Co and Fe$_{20}$Ni$_{80}$ layers are 139 and 149 Oe, respectively. The dispersive relations and the imaginary permeability dependence on external dc field $H$ were studied. An upward trend of permeability spectra peak value of the optical mode FMR is closely related to the out-of-phase precession of the magnetic moments. When the applied field $H$ increases gradually from 0 to 1700 Oe, there is a critical field $H_{crt}$ above which the peak value begin to decrease. Such a variation of peak value can be regulated via changing the layer thickness. It is thought that the changing of the layer thickness affects the equivalent exchange field. When Fe$_{20}$Ni$_{80}$ layer thickness rises from 33.4 to 66.8 nm, $H_{crt}$ drop off from 1600 Oe to 250 Oe, which can be qualitatively considered as a result of reduction of the effect interlayer exchange coupling field.
\begin{acknowledgments}
This work is supported by the National Basic Research Program of China (No. 2012CB933101), National Natural Science Foundation of China (NSFC) (Nos. 51471080, 11674143,
51371093, 51871117) and Program for Changjiang Scholars and Innovative Research Team in University(No.IRT-16R35).
\end{acknowledgments}

\nocite{*}

\bibliography{references}

\end{document}